\documentclass{birkjour}
\usepackage{graphicx}
\usepackage{caption}
\usepackage{subcaption}
\usepackage{mathrsfs}
\usepackage{color}
\usepackage[colorlinks=true,allcolors=blue]{hyperref}
\usepackage[utf8]{inputenc}
\usepackage{float}
 \newtheorem{thm}{Theorem}[section]

 \theoremstyle{definition}
 \newtheorem{defn}[thm]{Definition}
 \theoremstyle{remark}
 
 \newtheorem*{ex}{Example}
 \numberwithin{equation}{section}

\newcommand{\ket}[1]{|{#1}\rangle} %builds a ket (| >)
\newcommand{\bra}[1]{\langle{#1}|} %builds a bra (< |)
\newcommand{\grade}[1]{\left\langle{#1}\right\rangle} %buils grade (< >) %operator
\newcommand{\bracket}[2]{\left\langle{#1}|{#2}\right\rangle} %buils a bracket (<|>)

\begin{document}

%-------------------------------------------------------------------------
% editorial commands: to be inserted by the editorial office
%
%\firstpage{1} \volume{228} \Copyrightyear{2004} \DOI{003-0001}
%
%
%\firstpage{1}

\title[TSS Revisited: a Geometric Algebra Approach]{Two-State Quantum Systems Revisited:\\
 a Geometric Algebra Approach}

%----------Author 1
\author{Pedro Amao}

\address{%
Departamento de Ciencias, Secci\'on F\'isica\\
Pontificia Universidad Cat\'olica del Per\'u\\
Av. Universitaria 1801, San Miguel, Lima 32, Per\'u.}

\email{pedro.amao@pucp.edu.pe}

%\thanks{This work \ldots}
%----------Author 2
\author{Hern\'an Castillo}
\address{%
Departamento de Ciencias, Secci\'on F\'isica\\
Pontificia Universidad Cat\'olica del Per\'u\\
Av. Universitaria 1801, San Miguel, Lima 32, Per\'u.}

\email{hcastil@pucp.edu.pe}

\subjclass{Primary 81V45, Secondary 15A66} 

\keywords{Clifford algebras, Geometric algebra, Two-state quantum systems}

\date{\today}

\begin{abstract}
   We revisit the topic of two-state quantum systems using Geometric Algebra (GA) in three dimensions $\mathcal G_3$.
In this description, both the quantum states and  Hermitian operators are written as elements of $\mathcal G_3$.
By writing the quantum states as elements of the minimal left ideals of this algebra, we compute the energy eigenvalues and eigenvectors for the Hamiltonian of an arbitrary two-state system.
The geometric interpretation of the Hermitian operators enables us to introduce an algebraic method to diagonalize these operators in GA.
We then use this approach to revisit the problem of a spin-$1/2$ particle interacting with an external arbitrary constant magnetic field, obtaining the same results as in the conventional theory.
However, GA reveals the underlying geometry of these systems, which reduces to the Larmor precession in an arbitrary plane of $\mathcal G_3$.
\end{abstract}

%%% ----------------------------------------------------------------------
\maketitle
%%% ----------------------------------------------------------------------
%\tableofcontents
%%%%%%%%%%%%%%%%%%%%%%%%%%%%%%%%%%%%%%%%%%%%%%%%%%%%%%%%%%%%%%%%%%%%%%%%%%%%%%%%
\section{Introduction}
Two-state systems  (TSS) are ubiquitous in quantum theory.
One of the reasons for this is because many quantum systems can be considered (or approximated) to exist as a quantum superposition of only two distinguishable possible states.
Hence, their study can be greatly simplified because they can be described simply by a two-dimensional complex Hilbert space.
Alternatively, TSS can  also be studied using Clifford algebras, since the minimal left ideals of  a Clifford algebra are also appropriate for describing the spinor spaces and superpositions of quantum theory \cite{Hiley2012192}.
In mathematics, Clifford algebras have been extensively studied
\cite{lounesto2001clifford,vaz2016introduction}, and can be
considered as an associative algebra over a field.
The Clifford algebra over the field of real numbers  is best known in physics and
engineering   as  Geometric Algebra (GA) \cite{Zou2009147}.
Electromagnetic
fields \cite{Arthur2011, Dressel20151}, relativistic quantum
mechanics \cite{Hestenes2003691}, gravity gauge
theories \cite{Lasenby1998487}, and gravitational waves \cite{Lasenby2019} are all among the applications of GA to physics.

Quantum superposition in a TSS  is conventionally represented by the state vector $\ket\psi$ in the two dimensional complex Hilbert space $\mathcal H_2$.
The matrix representation of $\ket\psi$ is a two-component column matrix with complex entries known as a Pauli spinor,  which is an element of the spinor space $\mathbb C^2$.
In Clifford  algebras, $\mathbb C^2$ is isomorphic to the subspace of minimal left ideals of $\mathcal Cl_{3,0}$ \cite{Hiley2012192}.
Although the minimal left ideals are adequate for representing quantum superposition, in GA, Pauli spinors are usually mapped to the elements  of the even subalgebra $\mathcal G_3^+\subset\mathcal G_3$  \cite{Dargys2017241,doran_lasenby_2003,Francis2005383}.
We will show the advantages of using minimal left ideals instead of the elements of $\mathcal G_3^+$,
revisiting TSS, by representing the Pauli spinors as minimal left ideals of the GA in three dimensions.
To not get lost in this Babel of spinors, we adopt here the classification discussed in \cite{vaz2016introduction}, where the Pauli spinors $\ket\psi$ are classified as classical spinors, the minimal left ideals as algebraic spinors $\Psi$, and the elements of $\mathcal G_3^+$  as operator spinors $\psi_+$.

On the other hand, operators  such as spin are  conventionally represented by Hermitian or self-adjoint operators acting on classical spinors.
However, in GA these operators are in general elements of $\mathcal G_3$ acting on operator spinors.
For example, the action of the operator  on the classical spinor $(\hat H\ket\psi)$ results in another classical spinor, say $\ket\phi$.
By contrast, if we represent classical spinors by operator spinors, this  no longer holds.
The reason is that the operator spinors are not  invariant under left multiplication by the elements of $\mathcal G_3$, i.e.
for $H\in\mathcal G_3$, $H\psi_+\notin\mathcal G_3^+$.
One way to resolve this issue is by right multiplying $H\psi_+$ by one of the generators, say $\mathbf e_j\in\mathcal G_3$ \cite{doran_lasenby_2003}.
Then the resulting multivector  $H\psi_+\mathbf e_j$ remains in $\mathcal G_3^+$.
The approach discussed in this paper  does not have this limitation.

To our knowledge, algebraic spinors have already been used to represent classical spinors in GA \cite{Baylis2010517,McKenzie2015}.
For instance, by using algebraic spinors, it is possible to identify the eigenstates of the spin operator $\hat S_z$  with  two orthonormal elements of the minimal left ideal of $\mathcal G_3$.
These elements can be identified with the two poles of the Bloch sphere.
Another advantage is that the imaginary unit $\sqrt{-1}$  of conventional theory maps to the normalized sum of the three orthonormal bivectors of $\mathcal G_3$.
Additionally, the complex probability amplitudes of the TSS are mapped to the elements of the center of $\mathcal G_3$.
Hence, they commute with all the elements of the algebra, in the same way as complex numbers commute with the elements of $\mathcal H_2$.

We begin this study in Section~\ref{GA Review}, giving a short review of GA in an arbitrary number of dimensions, highlighting the three-dimensional case, which is fundamental for studying TSS.
Section~\ref{Quaternions in GA} treats the geometric interpretation of quaternions in GA.
Section~\ref{TSSGA} is the main part of the paper, dedicated to the study of TSS in GA.
We start by writing the states and operators of standard quantum mechanics in terms of the elements of $\mathcal G_3$.
In Section \ref{Spin Op} we discuss the algebra of spin operators in GA. Section~\ref{EigenvaluesEigenvectors} is dedicated to the application of GA to compute the eigenvalues and eigenvectors of an arbitrary Hermitian operator.
In Section~\ref{Spinonehalf}, we revisit the classical problem of spin-$1/2$ particle interacting with an arbitrary constant external magnetic field.
In  Section~\ref{Conclusions} we summarize the paper and then draw some conclusions.
In the Appendix, we review TSS in Hilbert space.

Throughout this paper, we will denote the elements of the field $\mathbb R$ by lowercase letters,  the generators of $\mathcal G_3$ in boldface $\{\mathbf e_i\}$, and their geometric product as $\mathbf e_l\mathbf e_m\ldots=\mathbf e_{lm\ldots}$.
The pseudoscalar of $\mathcal G_3$ will be written as $\mathbf i=\mathbf e_{123}$, and general multivectors of $\mathcal G_3$ in capitals.

%%%%%%%%%%%%%%%%%%%%%%%%%%%%%%%%%%%%%%%%%%%%%%%%%%%%%%%%%%%%%%%%%%%%%%%%%%%%%%%% 
\section{A Short Review of  Geometric Algebra}\label{GA Review}
%%%%%%%%%%%%%%%%%%%%%%%%%%%%%%%%%%%%%%%%%%%%%%%%%%%%%%%%%%%%%%%%%%%%%%%%%%%%%%%%
Geometric Algebra is a non-Abelian unital associative algebra over the field of real numbers.
The  elements of GA represent geometric objects, such as points, oriented lines, planes, and volumes.
Moreover, the algebraic operations between them can be interpreted as geometric operations such as rotations, reflections, and projections, as well as others.
\subsection{Real Quadratic Space}
Let  $\mathbb R^{p,q}$ be a real vector space  with orthonormal basis $\{\mathbf e_i\}$,  $(i=1,2\ldots p+q=n)$, such that
\begin{align}
    \mathbf e_i^2&=+1,\quad \text{for} \quad 1\leq i\leq p\\
    \mathbf e_i^2&=-1, \quad \text{for}\quad p+1\leq i\leq n.
\end{align}
Hence, if $x_i\in \mathbb R$, then any vector $\mathbf x\in\mathbb R^{p,q} $ can be written as
\begin{equation}
    \mathbf x=\underbrace{x_1\mathbf e_1+x_2\mathbf e_2+\ldots+x_p\mathbf e_p}_{\text{$p$ vectors}}+\underbrace{x_{p+1}\mathbf e_{p+1}+x_{p+2}\mathbf e_{p+2}+\ldots+x_{p+q}\mathbf e_{p+q}.}_{\text{$q$ vectors}}
\end{equation}  
The inner product $\mathbf x\cdot\mathbf x$ results in the quadratic form
\begin{equation}
\label{quad}
Q(\mathbf x) =\sum_{i=1}^p x_i^2-\sum_{i=p+1}^{p+q} x_i^2.
\end{equation}
This equation contains $p$ positive  and $q$ negative terms, so $(p,q)$ is known as the signature of $Q(\mathbf x)$.
The vector space $\mathbb R^{p,q}$ together with the quadratic form \eqref{quad} is called a non-degenerate real quadratic space.
\subsection{Definition of Geometric Algebra}
Among the several definitions of the Clifford algebras \cite[Ch. 14]{lounesto2001clifford}, we use here the definition by generators and relations.
This definition is suitable for the non-degenerate real quadratic spaces $\mathbb R^{p,q}$  discussed above to define  geometric algebra as
\begin{defn}
A unital associative algebra over $\mathbb R$ containing  $\mathbb R$ and $\mathbb R^{p,q}$ as distinct subspaces is the geometric algebra $\mathcal G_{p,q}$ if the following three conditions are satisfied \begin{itemize}
    \item [(i)] For $\mathbf x\in \mathbb R^{p,q}$, the geometric product denoted by $\mathbf x\mathbf x$ is equal to the quadratic form $Q(\mathbf x)$;
    \item [(ii)]  $\mathbb R^{p,q}$ generates $\mathcal G_{p,q}$ as an algebra over $\mathbb R$;
    \item [(iii)] $\mathcal G_{p,q}$ is not generated by any proper subspace of $\mathbb R^{p,q}$.
    \end{itemize}
      \end{defn}
\subsection{Geometric Algebra in Three Dimensions $(\mathcal G_3)$}
Here we use the above definition for the Euclidean space (signature  $(3,0))$ with the orthonormal basis
 $\{\mathbf e_1,\mathbf e_2,\mathbf e_3\}$  of $\mathbb R^{3,0}=\mathbb R^3$.
Therefore, condition (i) becomes
      \begin{align}
        &\mathbf e_i^2=+1\quad\text{for}\quad i=1,2,3,\\
        &\mathbf e_i\mathbf e_j=-  \mathbf e_j\mathbf e_i\quad \text{for}\quad i\neq j,\quad i,j=1,2,3.\label{eq:basis_anticom}
      \end{align}
      Condition (ii) implies that $\{\mathbf e_1,\mathbf e_2,\mathbf e_3\}$ generates a basis of  $\mathcal G_{3,0}=\mathcal G_3$ 
      \begin{equation}
        \label{eq:spanG3}
        \{1,\mathbf e_1,\mathbf e_2,\mathbf e_3,\mathbf e_{23},\mathbf e_{31},\mathbf e_{12},\mathbf e_{123}\},
      \end{equation}
which  corresponds to a basis for the exterior algebra $\bigwedge\mathbb R^3$
\begin{equation}
          \label{eq:spanR3}
          \{1,\mathbf e_1,\mathbf e_2,\mathbf e_3,\mathbf e_2\wedge\mathbf e_3,\mathbf e_3\wedge\mathbf e_1,\mathbf e_1\wedge\mathbf e_2,\mathbf e_1\wedge\mathbf e_2\wedge\mathbf e_3\}.
        \end{equation}
      The correspondence between \eqref{eq:spanG3} and \eqref{eq:spanR3} induces the decomposition
      \begin{equation}
        \label{eq:decomposition}
        \mathcal G_3=\mathbb R\oplus\mathbb R^3\oplus\bigwedge^2\mathbb R^3\oplus\bigwedge^3\mathbb R^3,
      \end{equation}
which introduces a multivector structure into $\mathcal G_3$, see \cite{lounesto2001clifford} for more details.
Note that (iii) is not needed here because $(p-q)\text{ mod }4\neq 1$ \cite{Lounesto1987,lounesto2001clifford}.

On the other hand, writing $1=\mathbf e_0$, $\mathbf i=\mathbf  e_{123}$ in \eqref{eq:spanG3}, and using the Hodge dual $A=\star \mathbf x=\mathbf{ix}$ for $\mathbf x\in \mathcal G_3$ and $A\in \bigwedge^2\mathbb R^3$,  we can write any  multivector $M\in\mathcal G_3$  as a sum: 
\begin{equation}
\label{multivector}
     M=\sum_{j=0}^{3}\lambda_j\mathbf e_j+ \mathbf i\sum_{j=0}^{3}\delta_j\mathbf e_j, \quad\text{where }\lambda, \delta\in\mathbb R.
       \end{equation}
       Note that in the first sum, there are one scalar and three vectors, whereas in the last one, there are one pseudoscalar and three bivectors.
Hence \eqref{multivector} exhausts all the elements of $\mathcal G_3$.

It is worth mentioning that using the pseudoscalar $\mathbf i\in \mathcal G_3$, we can write the geometric product of two arbitrary basis elements of $ \mathcal G_3$ as
\begin{equation}
  \label{eq:elm}
  \mathbf e_{lm}=\delta_{lm}+\epsilon_{lmn}\mathbf i\mathbf e_n,\quad\text{for }l,m,n=1,2,3,
\end{equation}
where $\delta_{lm}$ is the Kronecker delta function and $\epsilon_{lmn}$ is the Levi-Civita symbol.
Writing $\mathbf e_{ml}=\delta_{ml}+\epsilon_{mln}\mathbf i\mathbf e_n$ and subtracting this from \eqref{eq:elm} we have the useful relation
\begin{equation}
  \label{eq:elmcomm}
    \mathbf e_{lm}-  \mathbf e_{ml}=2\epsilon_{lmn}\mathbf i\mathbf e_n,\quad\text{for }l,m,n=1,2,3.
  \end{equation}
  
Lastly, we recall that an element $f\in \mathcal G_3$ is idempotent if $f^2=f$.
Two idempotents $f_1$ and $f_2$ are orthogonal if $f_1f_2=f_2f_1=0$.
An idempotent $f$ is primitive if $f\neq f_1+f_2$ \cite{Lounesto1987}.
\section{Geometric  Algebra of Quaternions}
   \label{Quaternions in GA}
The main achievement of Hamilton was to generalize the complex numbers to three dimensions, leading him to discover his quaternion algebra.
In this algebra, any quaternion can be written as
\begin{equation}
  \label{eq:quaternion}
  q=q_0+q_1i+q_2j+q_3k,
\end{equation}
where the coefficients $q_0\ldots q_3$  are real numbers, and $i$, $j$, and $k$ are imaginary units satisfying $i^2=j^2=k^2=ijk=-1$.

The quaternion algebra is isomorphic to the even subalgebra  $\mathcal G_3^+\subset\mathcal G_3$.
Hence, we have the correspondences: $i\rightarrow -\textbf e_{23}$, $j\rightarrow -\textbf e_{31}$ and $k\rightarrow -\textbf e_{12}$  \cite{lounesto2001clifford}.
Therefore \eqref{eq:quaternion} can be considered as the representation
of the even multivector
\begin{equation}
  \label{quaternion}
 \mathbf q= q_0-q_1\mathbf e_{23} -q_2\mathbf e_{31}-q_3\mathbf e_{12},
\end{equation}
which can be written compactly using the   pseudoscalar $\mathbf i$:
\begin{equation}
  \label{unnorm_quat}
\mathbf q=q_0-\mathbf{in},
\end{equation}
where $\mathbf  n=q_1\mathbf e_1+q_2\mathbf e_2+q_3\mathbf e_3$.

Let  $\hat{\mathbf n}=\mathbf n/|\mathbf n|$ be a unit vector parallel to $\mathbf n$ with $|\mathbf n|^2=q_1^2+q_2^2+q_3^2$  and  $|\mathbf q|^2=q_0^2+|\mathbf n|^2$.
Now dividing \eqref{unnorm_quat} by $|\mathbf q|\neq 0$ we have  the unit multivector 
\begin{equation}
  \label{eq:unit_quat}
  \hat{\mathbf q}=\frac{q_0}{|\mathbf q|}-\mathbf{i}\hat{\mathbf n}\frac{|\mathbf n|}{|\mathbf q|}.
\end{equation}
Setting $\cos\frac{\alpha}{2}=\frac{q_0}{|\mathbf q|}$ and  $\sin\frac{\alpha}{2}=\frac{|\mathbf n|}{|\mathbf q|}$, we get
\begin{equation}
    \hat{\mathbf q}=\cos\frac{\alpha}{2}-\mathbf i\hat{\mathbf n}\sin\frac{\alpha}{2},
      \end{equation}
      hence the  multivector \eqref{unnorm_quat} in polar and exponential  form is
      \begin{equation}
        \label{eq:pol-exp}
        \mathbf q=|\mathbf q|\hat{\mathbf q}=|\mathbf q|\left(\cos\frac{\alpha}{2}-\mathbf i\hat{\mathbf n}\sin\frac{\alpha}{2}\right)=|\mathbf q|\exp\left(-\mathbf i\hat{\mathbf n}\frac{\alpha}{2}\right).
      \end{equation}      
      This result   generalizes  complex numbers to three dimensions in GA as we show next.
First, we note that 
      \begin{equation}
        ( \mathbf i\hat{\mathbf n})^2=\mathbf i\hat{\mathbf n}\mathbf i\hat{\mathbf n}=\mathbf i^2{\hat{\mathbf n}}^2=-1,    
      \end{equation}
    therefore  $\mathbf i\hat{\mathbf n}$ is a unit bivector.
In  order to have a geometric picture of  $\mathbf i\hat{\mathbf n}$, let us write it in explicit form
      \begin{equation}
  \label{eq:planes}
  \mathbf i\hat{\mathbf n}=\frac{1}{\sqrt{q_1^2+q_2^2+q_3^2}}\left(q_1\boldsymbol{e}_{2}\wedge\boldsymbol{e}_3+ q_2\boldsymbol{e}_{3}\wedge\boldsymbol{e}_1+q_3\boldsymbol{e}_1\wedge\boldsymbol{e}_2\right).
\end{equation}
This bivector can have an arbitrary orientation, depending on the coefficients $q_1,q_2$ and $q_3$.
For the sake of simplicity, let $q_1=q_2=q_3=1$, and using the associative property  of the wedge product, we verify that ~\eqref{eq:planes} can be factorized in any of the three bivectors
\begin{align}
  P_1= &\frac{1}{\sqrt 3} (\boldsymbol{e}_{2}-\boldsymbol{e}_{1})\wedge (\boldsymbol e_3-\boldsymbol e_1),\nonumber\\
  P_2= &\frac{1}{\sqrt 3} (\boldsymbol{e}_{2}-\boldsymbol{e}_{1})\wedge (\boldsymbol e_3-\boldsymbol e_2),\\
    P_3= &\frac{1}{\sqrt 3}  (\boldsymbol{e}_{3}-\boldsymbol{e}_{1})\wedge (\boldsymbol e_3-\boldsymbol e_2),\nonumber
\end{align}
which means that $\mathbf i\hat{\mathbf n} =P_1=P_2=P_3$.
This follows because bivectors are equal if they have the same area and orientation.

\begin{figure}
        \includegraphics[scale=.85]{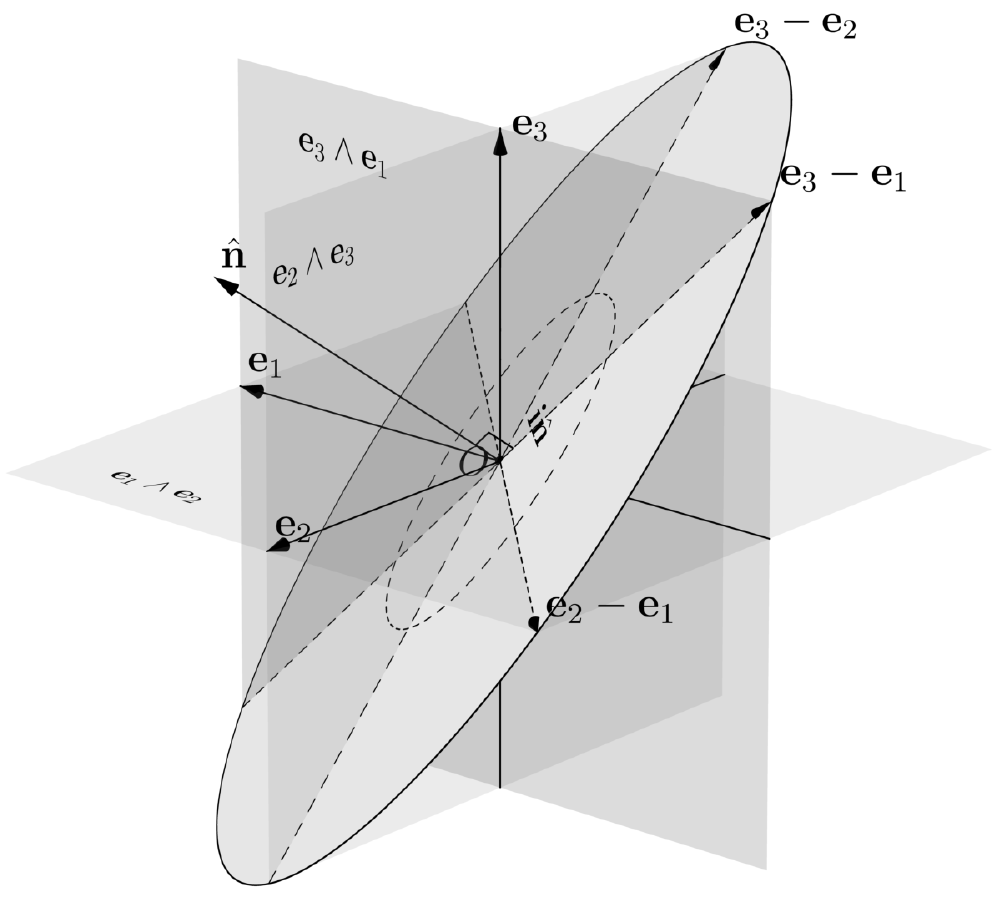}
        \caption{The unit bivector $\mathbf i\hat{\mathbf n}$ (dashed circle) is  the normalized sum of the three orthonormal planes $\mathbf e_1\wedge \mathbf e_2,  \mathbf e_2\wedge \mathbf e_3$ and  $\mathbf e_3\wedge \mathbf e_1$.}
        \label{fig:3dplane}
      \end{figure}

In Figure~\ref{fig:3dplane} we show the geometric representation of the bivector $\mathbf i \hat{\mathbf n}$.
Note that  $P_1, P_2$ and $P_3$ are generated by the vectors $\boldsymbol{e}_{2}-\boldsymbol{e}_{1}$, $\boldsymbol e_3-\boldsymbol e_1$ and $\boldsymbol{e}_{3}-\boldsymbol{e}_{2}$.
In addition, these vectors are in the intersections of the unnormalized plane $\mathbf i {\mathbf n}$ (great circle)
with the three orthogonal planes  $ \mathbf e_{23}, \mathbf e_{31}$ and $\mathbf e_{12}$.
Finally,  due to Hodge duality, $\mathbf i\hat{\mathbf n}$ is normal to the unit vector $\hat{\mathbf n}$.
      
      Overall, the above analysis shows that the bivector $\mathbf i\hat{\mathbf n}$ plays the role of the imaginary unit in three dimensions. This means that instead of having the  three imaginary units, $i,j$ and $k$, in GA we have the normalized sum of all three of them.
Therefore, the geometric representation of the pure quaternion $\frac{1}{\sqrt 3}(i+j+k)$ is an oriented  unit plane normal to the vector $\hat{\mathbf n}=\frac{1}{\sqrt 3}(\mathbf e_1+\mathbf e_2+\mathbf e_3)$.

      Having a geometric interpretation of the unit bivector, we see that \eqref{eq:pol-exp} represents an unnormalized rotor (operator spinor) producing counterclockwise rotations and dilations in the plane $\mathbf i\hat{\mathbf n}$.
This geometric interpretation is more transparent and intuitive than in the quaternion algebra $\mathbb H$, where the imaginary part of \eqref{eq:quaternion} $\Im(q)$ is interpreted as a vector of $\mathbb R^3$ having  negative square \cite{Morais20141}.

%%%%%%%%%%%%%%%%%%%%%%%%%%%%%%%%%%%%%%%%%%%%%%%%%%%%%%%%%%%%%%%%%%%%%%%%%%%%%%%%
\section{Two-State Quantum Systems in  Geometric Algebra}
\label{TSSGA}
%%%%%%%%%%%%%%%%%%%%%%%%%%%%%%%%%%%%%%%%%%%%%%%%%%%%%%%%%%%%%%%%%%%%%%%%%%%%%%%%
\subsection{States and Operators}
In GA the classical spinor $\ket \psi=c_+\ket + +c_-\ket -$ (see the Appendix), can be considered as the representation of an element of the minimal left ideal $S$ of $\mathcal G_3$.
The elements of $S$ are of the form $Af$ where $ A\in\mathcal G_3^+$ and $f$ is a primitive idempotent.
Therefore,  $\ket\psi$ maps to the algebraic spinor $\Psi$
\begin{equation}
    \label{eq:aspinor}
  \ket\psi\leftrightarrow\Psi=\psi_+ f,
\end{equation}
where $\psi_+$ can be written in terms of the elements of $\mathcal G_3^+$
\begin{equation}
\psi_+=a_0+a_1\mathbf e_{23} +a_2\mathbf e_{31}+a_3\mathbf e_{12}.
\end{equation}
Therefore, according to \eqref{quaternion} $\psi_+$ is a quaternion.
Furthermore,  due to the normalization condition of classical spinors $(\bracket{\psi}{\psi}=1)$, we require that $\psi_+\tilde\psi_+=\sum_{i=0}^3a_i^2=1$, hence $\psi_+$ must be a unit quaternion or rotor.

We now explain why \eqref{eq:aspinor} corresponds precisely to the classical spinor $\ket\psi$.
First, note that $\psi_+$ can be written as
\begin{equation}
  \psi_+  =(a_0+a_3\mathbf e_{12})+ (-a_2-a_1\mathbf e_{12})\boldsymbol e_{13}.
\end{equation}
Now we select the $z$ axis arbitrarily as the quantization direction, meaning that $f=\frac{1}{2}(1+\mathbf e_3)$.
Therefore
\begin{equation}
  \label{eq:spinortss}
  \Psi =(a_0+a_3\mathbf i)f +(-a_2+a_1\mathbf i)\mathbf e_1f,
\end{equation}
were we have used the ``Pacwoman property" $\mathbf e_3f=f$ \cite{baylis2004electrodynamics}.
Comparing \eqref{eq:spinortss} with \eqref{two_state} we have the correspondences \cite{vaz2016introduction}
\begin{equation}
  \label{eq:baseequiv}
 \begin{split}
 &c_+\leftrightarrow \mathfrak c_+=a_0+a_3\mathbf i,\quad c_-\leftrightarrow\mathfrak c_-= -a_2+a_1\mathbf i\\
 &\ket{+}\leftrightarrow \epsilon_+=f,\quad \ket{-}\leftrightarrow\epsilon_- =\mathbf e_1f  \\\
   &\ket\psi\leftrightarrow\Psi=\mathfrak c_+\epsilon_++\mathfrak c_-\epsilon_-.
 \end{split}
 \end{equation}
 Note that $\mathfrak c_+,\mathfrak c_-\in \text{Cen}(\mathcal G_3)$, hence they commute with all the multivectors of $\mathcal G_3$.
Additionally, we also need the  involutions
 \begin{equation}
  \label{eq:conjbaseequiv}
 \begin{split}
 &c_+^*\leftrightarrow \tilde{\mathfrak c}_+=a_0-a_3\mathbf i,\quad c_-^*\leftrightarrow\tilde{\mathfrak c}_-= -a_2-a_1\mathbf i\\
 &\bra{+}\leftrightarrow \tilde{\epsilon}_+=\epsilon_+,\quad \bra{-}\leftrightarrow\tilde{\epsilon}_- =f\mathbf e_1  \\\
   &\bra\psi\leftrightarrow\tilde\Psi=\tilde{\mathfrak c}_+\epsilon_++\tilde{\mathfrak c}_-\tilde{\epsilon}_-,
 \end{split}
 \end{equation}
 where ${}^\ast$ and ${}^\sim$ are the complex conjugation and reversion, respectively.
 
Finally, the  quantum inner product of the two orthonormal states $\ket +$ and $\ket -$  of the Hilbert space are written in $\mathcal G_3$ as follows
\begin{equation}  
\begin{split}
  \langle +|+\rangle \leftrightarrow 2\grade{\tilde\epsilon_+\epsilon_+}_0 =1,\\
    \langle -|-\rangle \leftrightarrow 2\grade{\tilde\epsilon_-\epsilon_-}_0 =1,\\
  \langle +|-\rangle \leftrightarrow 2\grade{\tilde\epsilon_+\epsilon_-}_0 =0,\\
   \langle -|+\rangle \leftrightarrow 2\grade{\tilde\epsilon_-\epsilon_+}_0 =0.
\end{split}
\end{equation}

To continue our discussion, we will see how to write the operators for a TSS in $\mathcal G_3$.
It is well known that the Hamiltonian of a TSS is represented by a linear combination of the Pauli and identity matrices (see the Appendix).
In the same way, by making use of the basis \eqref{eq:spanG3} we claim that any operator in $\mathcal G_3$ can be written  as the superposition
\begin{equation}
      \label{eq:HamMult}
     H=H_++H_-,
\end{equation}
where
\begin{align}
  H_+&=\sum_{j=0}^{3}h_j^+\mathbf e_j,\quad\text{where }  h_j^+\in\mathbb R,\label{eq:HermMult}\\
  H_-&=\mathbf i\sum_{j=0}^{3}h_j^-\mathbf e_j,\quad\text{where } h_j^-\in\mathbb R.\label{eq:NonHermMult}
\end{align}
Note that  \eqref{herm} is the matrix representation of $H_+$, whereas the matrix representation of $H_-$ is
\begin{equation}
\label{eq:nh_operator}
     \hat H_-=i\begin{pmatrix}
 h_0^-+h_3^- & h_1^--ih_2^- \\
 h_1^-+ih_2^- & h_0^--h_3^- 
\end{pmatrix}.
\end{equation}
Note that this  matrix is anti-Hermitian, and is conventionally considered as devoid of physical meaning.
However,  operators such as 
\eqref{eq:nh_operator}  may have a real spectrum if they are symmetric under the   Parity--Time $(\mathscr{PT})$ transformation in non-Hermitian quantum mechanics (NHQM) \cite{PhysRevLett.80.5243}.
NHQM is outside the scope of this paper, therefore we will only consider  operators such as \eqref{eq:HermMult}.

%%%%%%%%%%%%%%%%%%%%%%%%%%%%%%%%%%%%%%%%%%%%%%%%%%%
\subsection{Geometric Algebra of Spin Operators}\label{Spin Op}
%%%%%%%%%%%%%%%%%%%%%%%%%%%%%%%%%%%%%%%%%%%%%%%%%%%

In quantum mechanics, the spin operators defined as $\hat{S}_i=\frac{\hbar}{2}\hat\sigma_i$ for $i=1,2,3$ obey the commutation relations
\begin{equation}
\label{qcom}
    [\hat S_i,\hat S_j]=i\hbar\epsilon_{ijk}\hat S_k.
\end{equation}
However, due to the isomorphism between the matrix algebra  $\text{Mat}(2,\mathbb C)$ and $\mathcal G_3$, we have the correspondences
\begin{equation}
\label{spinvectors}
    \hat S_i=\frac{1}{2}\hbar \hat \sigma_i\leftrightarrow \mathbf S_i=\frac{1}{2}\hbar \mathbf e_i,\quad\text{for }i=1,2,3.
      \end{equation}
      Therefore, in $\mathcal G_3$ the spin operators  $\hat{S}_i$ become the spin vectors $\mathbf S_i$.
With these equivalences in mind and writing the LHS of \eqref{qcom} in terms of the spin vectors, we have
      \begin{equation}
  \mathbf S_i\mathbf S_j-  \mathbf S_j\mathbf S_i= \frac{1}{4}\hbar^2(\mathbf e_{ij}-\mathbf e_{ji}).
\end{equation}
Substituting \eqref{eq:elmcomm} in the RHS of the above result, we arrive at the commutator
\begin{equation}
\label{ccom}
[\mathbf S_i,\mathbf S_j]=\hbar\epsilon_{ijk} \mathbf i\mathbf S_k,
\end{equation}
which is the GA version of the algebra of spin operators of quantum mechanics, where the role of the imaginary unit in \eqref{qcom}  has been taken by the pseudoscalar of $\mathcal G_3$ because \eqref{ccom} is a bivector identity.

It is well known that \eqref{qcom} extends to total $(\hat J)$ and orbital $(\hat L)$ angular momentum operators, respectively.
Therefore, \eqref{ccom} must extend to the corresponding total and orbital angular momentum vectors.
Since in GA the orthogonal components of these operators are all vector quantities, it is not surprising that they must obey the same commutation relations.
%%%%%%%%%%%%%%%%%%%%%%%%%%%%%%%%%%%%%%%%%%%%%%%%%%%%%%%%%%%%%%%%%%%%%%%%%%%%%%%%
\subsection{Eigenvalues and Eigenvectors}\label{EigenvaluesEigenvectors}
%%%%%%%%%%%%%%%%%%%%%%%%%%%%%%%%%%%%%%%%%%%%%%%%%%%%%%%%%%%%%%%%%%%%%%%%%%%%%%%%
One of the most important problems in quantum mechanics is to determine the energy eigenvalues and eigenvectors of the Hamiltonian
\begin{equation}
\label{HermitianOperator}
  \hat H = h_0\hat\sigma_0+h_1\hat \sigma_1+h_2\hat \sigma_2+h_3\hat \sigma_3=
  \begin{pmatrix}
 h_0+h_3 & h_1-ih_2 \\
 h_1+ih_2 & h_0-h_3 
\end{pmatrix},
\end{equation}
which is a self-adjoint or Hermitian matrix.

The standard method to solve this problem consists in finding the eigenvalues by solving the characteristic equation $\text{det}(\hat H-\lambda \hat \sigma_0)=0$.
 Then we use these eigenvalues to find the eigenvectors by solving the eigenvalue equations $\hat H\ket\psi=\lambda\ket\psi$.
The energy eigenvalues are the entries of the diagonal matrix 
\begin{equation}
\label{eq:diagonalmat}
\hat H_0=\hat R^{-1}\hat H\hat R,
\end{equation}
where $\hat R$ is an invertible matrix built from the eigenvectors of $\hat H$.

It is well known that if $\hat H$ is diagonalizable, then the eigenvalue problem reduces to the simple form
\begin{equation}
 \hat H_0\ket \pm=E_\pm\ket \pm,
\end{equation}
where $E_\pm$ and $\ket\pm$ are the eigenvalues and eigenvectors of the Hamiltonian $\hat H$, respectively.

Here, we develop a practical method to solve this problem in GA.
Let us start by recalling that  the operator $\hat H$ is the matrix representation of  the multivector  \eqref{eq:HermMult}
\begin{equation}
  \label{eq:SelfMult}
  H=h_0+\mathbf r=h_0+|\mathbf r|\hat{\mathbf r},
\end{equation}
where $|\mathbf r|=\sqrt{h_1^2+h_2^2+h_3^2}$,
$\hat {\mathbf r}$ is a unit vector parallel to $\mathbf r$, and we have written for simplicity $h_j^+=h_j$.
To continue, note that in Mat$(2,\mathbb C)$ the only diagonal matrix (apart from the identity matrix) is $\hat\sigma_3$.
Therefore, all diagonal matrices of Mat$(2,\mathbb C)$ must be  scalar multiples of $\hat\sigma_3$.

On the other hand,
$\mathbf r$ has an arbitrary direction in $\mathcal G_3$.
Hence, in analogy to \eqref{eq:diagonalmat}, we can always
make $\mathbf r$ parallel to $\mathbf e_3$ by rotating the multivector $H$:
\begin{equation}
  \label{eq:hamrot}
  H_0=\tilde R H R=h_0+|\mathbf r|\tilde R\hat{\mathbf r} R=h_0+|\mathbf r|\mathbf{e_3}.
\end{equation}
Consequently, the matrix representation of the above equation results in a
diagonal matrix
\begin{equation}
  \hat {H}_0=
    \begin{pmatrix}
 h_0+|\mathbf r|& 0\\
  0& h_0- |\mathbf r|
\end{pmatrix},
\end{equation}
and so we have diagonalized $\hat H$ just by rotating the vector part of $H$.

The operator spinor or rotor $R$ introduced in \eqref{eq:hamrot} can be written in terms of the coefficients $h_1$, $h_2$ and $h_3$ as
\begin{equation}
  \label{eq:diagrotor}
  R=R(\varphi)R(\theta)=\exp\left(-\mathbf{ie}_3\varphi/2\right)\exp\left(-\mathbf{ie}_2\theta/2\right),
\end{equation}
where the polar angles are given by
\begin{equation}
  \label{eq:theta}
  \theta=\text{tan}^{-1}\left(\frac{\sqrt{h_1^2+h_2^2}}{h_3}\right),
\end{equation}
\begin{equation}
  \label{eq:varphi}
  \varphi=\tan^{-1}(\frac{h_2}{h_1}).
\end{equation}

Having found  $H_0$ and due to the equivalences \eqref{eq:baseequiv}, the eigenvalue equations in GA can be written as
\begin{equation}
 \hat H_0\ket \pm=E_\pm\ket\pm\leftrightarrow H_0\epsilon_\pm=E_\pm\epsilon_\pm.
\end{equation}

On the order hand, in order to find the eigenvectors of $\hat H$,  note that the vector part of \eqref{eq:SelfMult} has a direction given by the unit vector $\hat{\mathbf r}=\sin\theta\cos\varphi\mathbf e_1+\sin\theta\sin\varphi\mathbf e_2+\cos\theta\mathbf e_3$.
Therefore, we may proceed in the same way that the eigenvectors of the spin operator in the arbitrary direction $\mathbf{\hat n}$  are found in quantum mechanics, see, for example, \cite[Ch. 3]{sakurai2011modern}.

Suppose that at $t=0$ the classical spinor \eqref{two_state} is in the up state, $\ket{\psi(t=0)}=\ket +$.
Then, using  \eqref{eq:baseequiv} yields $\mathfrak c_+=1$ and $\mathfrak c_-=0$, and hence $\Psi(t=0)=\Psi_+=\epsilon_+$.
Now we perform two successive rotations $R(\theta)$ and $R(\varphi)$ on $\Psi_+$:
\begin{equation}
  \label{eigenspinor+}
 \Psi_{\hat{\mathbf r}+}=R(\varphi)R(\theta)\Psi_+=\exp\left(-\mathbf{ie}_3\varphi/2\right)\exp\left(-\mathbf{ie}_2\theta/2\right)\epsilon_+.
\end{equation}
The algebraic spinor $ \Psi_{\hat{\mathbf r}+}$  corresponds to  one of the eigenvectors of $\hat H$.
The other  algebraic spinor $\Psi_{\hat{\mathbf r}-}$ can be found by rotating $\Psi_+$ again, but this time we change $R(\theta)$ to $R(\theta+\pi)$
\begin{equation}
   \label{eigenspinor-}
  \Psi_{\hat{\mathbf r}-}=R(\varphi)R(\theta+\pi)\Psi_+=\exp\left(-\mathbf{ie}_3\varphi/2\right)\exp\left(-\mathbf{ie}_2(\theta/2+\pi/2)\right)\epsilon_+.
\end{equation}
The algebraic spinors $ \Psi_{\hat{\mathbf r}+}$ and $ \Psi_{\hat{\mathbf r}-}$ are orthonormal as required.

With the above equivalences, the eigenvalue equations $H\Psi_{\hat{\mathbf r}\pm}=E_{\pm} \Psi_{\hat{\mathbf r}\pm}$ become
\begin{equation}
  \begin{split}
    H  \Psi_{\hat{\mathbf r}\pm}&=(h_0+|\mathbf r|\hat{\mathbf r}) \Psi_{\hat{\mathbf r}\pm}\\
    &=(h_0\pm|\mathbf r|) \Psi_{\hat{\mathbf r}\pm}.
  \end{split}
\end{equation}

%%%%%%%%%%%%%%%%%%%%%%%%%%%%%%%%%%%%%%%%%%%%%%%%%%%%%%%%%%%%%%%%%%%%%%%%%%%%%%%%
\begin{ex}
%%%%%%%%%%%%%%%%%%%%%%%%%%%%%%%%%%%%%%%%%%%%%%%%%%%%%%%%%%%%%%%%%%%%%%%%%%%%%%%%
Let $\hat H=a(\hat\sigma_1+\hat\sigma_3)$ be the Hamiltonian operator  for  a TLS, where $a$ is a real constant having the dimensions of energy. The problem is to find the corresponding energy eigenvalues and energy eigenvectors as linear combinations of the up and down states.

In order to solve this problem within $\mathcal G_3$, we begin by writing the operator $\hat H$ as $H =a(\mathbf e_1+\mathbf e_3)$.
Using Equations \eqref{eq:theta} and \eqref{eq:varphi} we have $\varphi=0$ and $\theta= \pi/4$. Hence the rotor which makes  $H$ parallel to $\mathbf e_3$ is given by \eqref{eq:diagrotor}
\begin{equation}
  R=\exp\left(-\mathbf{ie}_2\pi/8\right).
\end{equation}
Now, the rotated Hamiltonian becomes
\begin{equation}
  H_0=\tilde RHR=a\sqrt{2}\mathbf e_3.
\end{equation}
The matrix representation of this result reveals the two eigenvalues of $\hat H$:
\begin{equation}
  \hat H_0= \begin{pmatrix}
a\sqrt{2} & 0 \\
 0 & -a\sqrt{2}
\end{pmatrix}.
\end{equation}

On the other hand, using \eqref{eigenspinor+} and \eqref{eigenspinor-}, the algebraic spinors associated to $H$ are
\begin{align}
  \Psi_{\hat{\mathbf r}+}=&\exp\left(-\mathbf{ie}_2\pi/8\right)\epsilon_+,\\
  \Psi_{\hat{\mathbf r}-}=&\exp\left(-\mathbf{ie}_25\pi/8\right)\epsilon_+,
\end{align}
from which we can easily calculate their matrix representations, see Problem 1.10 of \cite{sakurai2011modern},
\begin{align}
  \ket{\Psi_{\hat{\mathbf r}+}}=&  \begin{pmatrix}
 \frac{\sqrt{2+\sqrt{2}}}{2} & 0 \\
 \frac{\sqrt{2-\sqrt{2}}}{2} & 0
\end{pmatrix}\\
   \ket{\Psi_{\hat{\mathbf r}-}}=&  \begin{pmatrix}
 -\frac{\sqrt{2-\sqrt{2}}}{2} & 0 \\
 \frac{\sqrt{2+\sqrt{2}}}{2} & 0
\end{pmatrix}.
\end{align}
\end{ex}
%%%%%%%%%%%%%%%%%%%%%%%%%%%%%%%%%%%%%%%%%%%%%%%%%%%%%%%%%%%%%%%%%%%%%%%%%%%%%%%%

\subsection{A Spin-$1/2$ Particle Interacting  with an External Magnetic Field}\label{Spinonehalf}
%%%%%%%%%%%%%%%%%%%%%%%%%%%%%%%%%%%%%%%%%%%%%%%%%%%%%%%%%%%%%%%%%%%%%%%%%%%%%%%%

In standard quantum mechanics, the Hamiltonian of a non-relativistic particle interacting with an external magnetic field $\mathbf B$ is given by 
\begin{equation}
\label{int_Ham}
 \hat H=-\frac{q\hbar}{2m}\boldsymbol\sigma\cdot \mathbf B,
\end{equation}
where $q$ is the charge of the particle, $m$ is its mass and $\hbar$ is the reduced Planck's constant.
Recalling that $\boldsymbol \sigma\cdot \mathbf B$ is just a short hand for $\hat\sigma_1B_1+\hat\sigma_2B_2+\hat\sigma_3B_3 $, we write \eqref{int_Ham} explicitly as
\begin{equation*}
 \hat H=-\frac{q\hbar}{2m}(B_1\hat\sigma_1+B_2\hat\sigma_2+B_3\hat\sigma_3).
 \end{equation*}
Note that this equation is similar to the matrix representation of the multivector \eqref{eq:HermMult} except for the term $h_0$, which  only changes  the reference level of the energy eigenvalues of \eqref{int_Ham}.
Therefore, in   $\mathcal G_3$ this Hamiltonian can be readily written down as
\begin{equation}
\label{Hamiltonian}
H=-\frac{q\hbar}{2m}\left(B_1\mathbf e_1+B_2\mathbf e_2+B_3\mathbf e_3\right)=-\frac{q\hbar}{2m}\mathbf B,
\end{equation}
representing an arbitrary vector of $\mathcal G_3$.

We now proceed to discuss the dynamics of the state vector wich is described by the time-dependent Schrödinger equation, (see the Appendix)
\begin{equation}
     \frac{\mathrm d\ket{\psi}}{\mathrm d t}=-\frac{1}{\hbar}(i\hat H)\ket{\psi}.
\end{equation}
This equation is straightforwardly translated to $\mathcal G_3$ as
\begin{equation}
  \label{eq:SchGA}
 \frac{\mathrm d\Psi}{\mathrm d t}=-\frac{1}{\hbar}(\mathbf i H)\Psi,
\end{equation}
where we have used \eqref{eq:aspinor}, replaced the imaginary unit $i$ by the pseudoscalar $\mathbf i$, and the Hamiltonian \eqref{int_Ham} by the multivector \eqref{Hamiltonian}.
Note that the product $(\mathbf iH)\Psi$ on the RHS of \eqref{eq:SchGA} will always remain in $\mathcal G_3^+f$ since $\Psi$ is a minimal left ideal.
Therefore, there is no need to multiply the RHS of \eqref{eq:SchGA}  by $\mathbf e_3$ in order to ensure that the resulting multivector will remain in $\mathcal G_3^+$, as discussed previously in \cite{doran_lasenby_2003}.

Now, if the vector $H$ is time independent (constant magnetic field), Equation~\eqref{eq:SchGA} can be easily integrated, yielding
\begin{equation}
\label{eq:ideal_evol}
\Psi(t)= \exp{\left(-\frac{\mathbf iHt}{\hbar}\right)}{\Psi}(0).
\end{equation}
Note that \eqref{evolution} is the matrix representation of \eqref{eq:ideal_evol} due to the isomorphism $G_3^+f\simeq \mathbb C^2$.
Hence, we have the mapping
\begin{equation}
  \label{eq:evolution}
\hat U\leftrightarrow U=\exp{\left(-\frac{\mathbf iHt}{\hbar}\right)}.
\end{equation}
Thus,  in $\mathcal G_3$ the evolution operator $\hat U$ is the matrix representation 
 of the rotor $U=\exp{(-\frac{\mathbf iHt}{\hbar})}\in\mathcal G_3^+$,
which is a rotor since the factor $\mathbf{i}H$ in its exponent is a bivector.
Writing  $U$ in terms of \eqref{Hamiltonian} gives us 
\begin{equation}
\label{spin_rotor}
U=\exp{\left(\mathbf i\hat{\mathbf B}\frac{\alpha}{2}\right)},
\end{equation}
where $\hat{\mathbf B}$ is a unit vector parallel to ${\mathbf B}$ and $\alpha=\frac{q|\mathbf B|t}{m}$.

%%%%%%%%%%%%%%%%%%%%%%%%%%%%%%%%%%%%%%%%%%%%%%%%%%% 
\subsection{Expectation Values}
%%%%%%%%%%%%%%%%%%%%%%%%%%%%%%%%%%%%%%%%%%%%%%%%%%%
Other quantities of interest in quantum mechanics are the expectation values of the Hermitian operator \eqref{HermitianOperator}, conventionally computed by
\begin{equation}
    \langle \hat H\rangle=\langle\psi| \hat H|\psi\rangle.
      \end{equation}
      Using the equivalences \eqref{eq:baseequiv}, the above definition  corresponds to taking the scalar part of the product $\tilde\Psi H\Psi$ and multiplying it by 2.
See, for example, \cite{lounesto2001clifford}
      \begin{equation}
        \label{eq:expect}
  \langle H\rangle=2\langle\tilde\Psi H\Psi\rangle_0.
\end{equation}

\begin{ex}[Spin Precession]
As an application of \eqref{eq:expect}, we will compute the expectation values of the spin  operators $\hat S_i=(\hbar/2)\hat\sigma_i$.
As we have seen in \eqref{spinvectors}, $\hat S_i$ corresponds to the spin vectors $\mathbf S_i=(\hbar/2)\mathbf e_i$ ($i=1,2,3$).

Recalling that $\Psi(t)$ is given by \eqref{eq:ideal_evol}, letting $\Psi(0)=\text{exp}(-\mathbf {ie}_2\theta/2)\epsilon_+$, and setting  $\hat{\mathbf B}=\mathbf e_3$ in \eqref{spin_rotor}, we have
\begin{equation}
  \Psi(t)=U\Psi(0)=\text{exp}(\mathbf{ie}_3\frac{\alpha}{2})\text{exp}(-\mathbf {ie}_2\frac{\theta}{2})\epsilon_+.
\end{equation}
Thus, the expectation value of $\mathbf S_1$ becomes
\begin{align}
  \langle \mathbf S_1\rangle&=\hbar\langle\tilde\Psi(t)\mathbf e_1\Psi(t)\rangle_0\\
                    &=\hbar\langle\epsilon_+\text{exp}(\mathbf{ie}_2\frac{\theta}{2})\text{exp}(-\mathbf {ie}_3\frac{\alpha}{2})\mathbf e_1\text{exp}(\mathbf{ie}_3\frac{\alpha}{2})\text{exp}(-\mathbf {ie}_2\frac{\theta}{2})\epsilon_+\rangle_0\\
                    &=\hbar\langle\sin(\theta)\cos(\omega t)\epsilon_++\cos(\theta)\cos(\omega t)\epsilon_+\epsilon_-+\sin(\omega t)\epsilon_+\mathbf e_2\epsilon_+\rangle_0\\
  &=\frac{\hbar}{2}\sin(\theta)\cos(\omega t),
\end{align}
since $\epsilon_+\epsilon_-=\epsilon_+\mathbf e_2\epsilon_+=0$ and $\langle\epsilon_+\rangle_0=1/2$.

In the same fashion, the expectation values of $\mathbf S_2$ and $\mathbf S_3$ are
\begin{align}
  \langle \mathbf S_2\rangle&=-\frac{\hbar}{2}\sin(\theta)\sin(\omega t),\\
    \langle \mathbf S_3\rangle&=\frac{\hbar}{2}\cos(\theta),
\end{align}
which describes the precession of $\langle\mathbf S\rangle$ in the plane $\mathbf{ie}_3$ at the Larmor  frequency $\omega=qB_3/m$.

Alternatively, we obtain all the components of the spin vector $\mathbf S$ easily using the operator spinor $\psi_+=\text{exp}(\mathbf{ie}_3\frac{\alpha}{2})\text{exp}(-\mathbf {ie}_2\frac{\theta}{2})$ acting on $\mathbf S_3$ as follows, see \cite{lounesto2001clifford} for more details.
\begin{equation}
  \mathbf S=\psi_+\mathbf S_3\tilde \psi_+=\frac{\hbar}{2}(\sin(\theta)\cos(\omega t)\mathbf e_1-\sin(\theta)\sin(\omega t)\mathbf e_2+\cos(\theta)\mathbf e_3).
\end{equation}
Therefore, the expectation values $\langle \mathbf S_i\rangle$ can be readily obtained as
\begin{equation}
  \langle \mathbf S_i\rangle=\mathbf S\cdot\mathbf e_i,\quad i=1,2,3.
\end{equation}
\end{ex}
%%%%%%%%%%%%%%%%%%%%%%%%%%%%%%%%%%%%%%%%%%%%%%%%%%%
\subsection{Probability}
%%%%%%%%%%%%%%%%%%%%%%%%%%%%%%%%%%%%%%%%%%%%%%%%%%%
Conventionally, if $\hat A$ is a Hermitian operator with a non-degenerate  discrete spectrum with eigenvalues $a_n$, then there is a unique eigenvector $\ket {u_n}$ associated to each eigenvalue $a_n$.
Since the set of $\ket{u_n}$ constitutes a basis, then $\ket{\psi(t)}=\sum_nc_n\ket{u_n}$.
Therefore, the probability of finding $a_n$ when $\hat A$ is measured is given by \cite{cohen1991quantum}
\begin{equation}
  \label{eq:convprob}
 P(a_n)=|c_n|^2=\langle u_n|\psi(t)\rangle\langle\psi(t)|u_n\rangle.
\end{equation}

Again, using the equivalences \eqref{eq:baseequiv}, and \eqref{eq:conjbaseequiv}, the translation of  \eqref{eq:convprob} to $\mathcal G_3$  yields
\begin{equation}
  \label{eq:gaprob}
  P(a_n)=2\langle\tilde u_n\Psi(t)\tilde\Psi(t)u_n\rangle_0.
\end{equation}

\begin{ex}[Rabi Formula]
In the preceding example, the direction of the magnetic field was along the positive $z$-axis.
Now, imagine that the spin  is initially in the eigenstate $\ket +$ of $\hat S_3$, and then we set the magnetic field in an arbitrary direction  $\mathbf B=B_1\mathbf e_1+B_2\mathbf e_2+B_3\mathbf e_3$.
We will now see how the probability $P_{+-}(t)$ of finding the spin in the eigenstate $\ket -$ of $\hat S_3$ at time $t$ is calculated in GA.

First, according to \eqref{two_state}, the state vector is given in terms of the basis  $\ket\pm$ of $\hat S_3$ by $\ket \psi=c_+\ket ++c_-\ket -$.
Now, if the spin is initially in the eigenstate $\ket +$ of $\hat S_3$, then $c_+=1$ and $c_-=0$; therefore, using  \eqref{eq:baseequiv}, the algebraic spinor at time $t=0$ becomes $\Psi(0)=\epsilon_+$.
Consequently, at time $t$ we have  
\begin{equation}
  \Psi(t)=U\Psi(0)=\exp\left(\frac{\mathbf i(B_1\mathbf e_1+B_2\mathbf e_2+B_3\mathbf e_3)}{\sqrt{B_1^2+B_2^2+B_3^2}}\frac{\alpha}{2}\right)\epsilon_+,
\end{equation}
where we have written explicitly the rotor \eqref{spin_rotor} in terms of the arbitrary field $\mathbf B$.

Having  $\Psi(t)$, now we can find the probability $P_{+-}(t)$ using the formula \eqref{eq:gaprob}:
\begin{equation}
  \label{eq:prob+-}
  P_{+-}(t)=2\langle\tilde \epsilon_-\Psi(t)\tilde\Psi(t)\epsilon_-\rangle_0
=2\langle\tilde \epsilon_-(U\epsilon_+\tilde U)\epsilon_-\rangle_0.
\end{equation}
Expanding the term in parentheses in the RHS of \eqref{eq:prob+-},  we have
\begin{equation}
  U\epsilon_+\tilde U=\frac{1}{2}\left(1+\mathbf u(t)\right),
\end{equation}
where $\mathbf u(t)$ is given by
\begin{equation}
    \label{eq:e3precess}
  \begin{split}
    \mathbf u(t)&=U\mathbf e_3\tilde U=\frac{1}{|\mathbf B|}\left(B_1\cos\theta(1-\cos\alpha)+B_2\sin\alpha\right)\mathbf e_1\\
    & +\frac{1}{|\mathbf B|}\left(B_2\cos\theta(1-\cos\alpha)-B_1\sin\alpha\right)\mathbf e_2+\left(\cos^2\theta+\sin^2\theta\cos\alpha\right)\mathbf e_3.
      \end{split}
\end{equation}
Here $\theta$ is the angle between the field $\mathbf B$ and  $\mathbf e_3$, see Figure~\ref{fig:spinprecess}.

Using this result, the probability $P_{+-}(t)$  reduces to
\begin{equation}
  P_{+-}(t)=2\langle \frac{1}{2}\left(1-\cos^2\theta-\sin^2\theta\cos\alpha\right)\epsilon_+\rangle_0.
\end{equation}
Finally, we can simplify this  expression by writing $1=\cos^2\theta+\sin^2\theta$ and $\alpha=\omega t$:
\begin{equation}
  \label{eq:rabiprobalt}
   P_{+-}(t)=\frac{1}{2} \sin^2\theta\left(1-\cos\omega t\right).
 \end{equation}
  Equation \eqref{eq:rabiprobalt} is best known as Rabi's formula. It gives us the probability of finding the system at time $t$ in the eigenstate $\epsilon_-$.
 
  The geometric interpretation of this result is shown in Figure~\ref{fig:spinprecess}.
If we define  $\mathbf S_u(t)=(\hbar/2)\mathbf u(t)$, then according to \eqref{eq:e3precess}, at time $t=0$, $\mathbf S_u(t)$ reduces to
  \begin{equation}
     \mathbf S_u(0)=\frac{\hbar}{2}\mathbf e_3.
   \end{equation}
Therefore, initially the spin vector is parallel to $\mathbf e_3$.
Then as times goes on, $\grade{\mathbf S_u(t)}$  precesses clockwise  in the plane $\mathbf i\hat{\mathbf B}$ at the Larmor frequency $\omega=\frac{q|\mathbf B|}{m}$.
See \cite[Ch, IV]{cohen1991quantum} for an alternative geometric interpretation of Rabi's formula.

 \begin{figure}
   \includegraphics[scale=1]{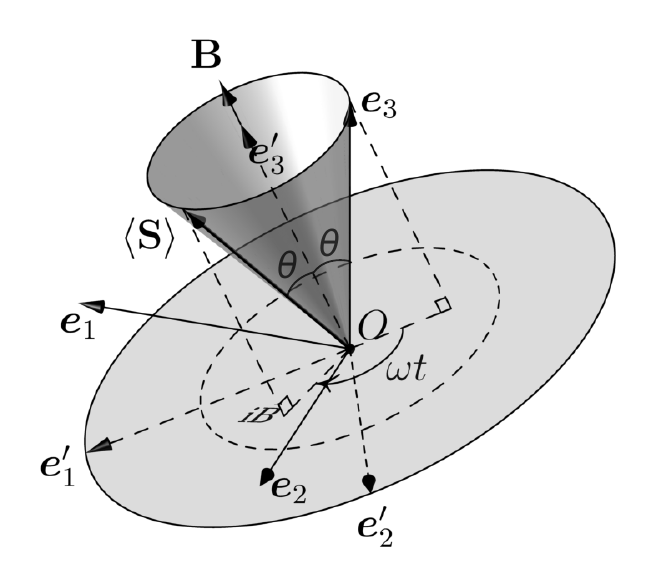}
   \caption{Geometric interpretation of spin precession for an arbitrary field  $\mathbf B$.}
        \label{fig:spinprecess}
\end{figure}

\end{ex}

\section{Summary and Conclusions}\label{Conclusions}
We have studied TSS  entirely within the geometric algebra of three dimensional space.
In this approach, the mathematical tools conventionally used to describe TSS, namely the complex numbers, linear vector spaces, eigenvalues, eigenvectors, etc., are unified into a single algebra.
This unification supports the claim that ``GA provides a unified language for the whole physics that is conceptually and computationally superior to alternative systems in every application domain" \cite{Hestenes2003104}.

Writing the Pauli spinors  and the self-adjoint operators of conventional quantum mechanics as elements of $\mathcal G_3$, we have computed the energy eigenvalues and eigenvectors of an arbitrary TSS.
Also, the geometric interpretation of the Hermitian operators of the TSS enables us to diagonalize them just by rotating their vector part.
In fact, we do not have to use \eqref{eq:hamrot} to find the correspondig diagonal matrix, because rotors preserve the norm of a vector, so it suffices to compute their norm and then multiply the result by $\mathbf e_3$.
This geometric interpretation is not present in the conventional theory.

We also revisited the problem of a spin-$1/2$ particle interacting with an external magnetic field, where the interaction Hamiltonian of this system has been interpreted as a vector of $\mathcal G_3$.
We have shown that the time evolution of the state vector reduces to the single side transformation law of algebraic spinors.
This also shows that the time evolution operator of quantum mechanics is represented as an operator spinor of $\mathcal G_3$.
We have also computed the expectation values and transition probabilities of this system, obtaining the same results as the conventional theory.
However, the algebraic approach reveals the underlying geometry of the probability transition $P_{+-}(t)$, which reduces to the Larmor precession in the arbitrary plane $\mathbf i\hat{\mathbf B}$.

Lastly, we have rewritten the algebra of spin operators \eqref{qcom} as the geometric algebra of the spin vectors \eqref{ccom} where the pseudoscalar of $\mathcal G_3$ plays the role of the imaginary unit $\sqrt{-1}$.
However, whereas in the conventional theory this imaginary unit is introduced \textit{ad hoc} to ensure that the commutator of the LHS of \eqref{qcom} results in a Hermitian matrix, the pseudoscalar $\mathbf i$ sends the spin vector $\mathbf S_k$, up to a multiplicative constant, to the spin bivector $\mathbf S_i\mathbf S_j$ via Hodge duality.
Therefore, the algebra of spin vectors \eqref{ccom} results from the anticommutativity of the geometric product.

\subsection*{Acknowledgment}
This work was supported by the Huiracocha grant from the Graduate School of the Pontificia Universidad Cat\'olica del Per\'u.

\section*{Appendix}
%%%%%%%%%%%%%%%%%%%%%%%%%%%%%%%%%%%%%%%%%%%%%%%%%%%%%%%%%%%%%%%%%%%%%%%%%%%%%%% 
\section*{Two-State Quantum Systems in a Hilbert Space}
\label{Appendix}
%%%%%%%%%%%%%%%%%%%%%%%%%%%%%%%%%%%%%%%%%%%%%%%%%%%%%%%%%%%%%%%%%%%%%%%%%%%%%%%%
Conventionally,  a two-state quantum system is  mathematically encoded using  a two-dimensional complex  Hilbert space $\mathcal H_2$.
In this space, the TSS, represented by the state vector or classical spinor  $\ket{\psi}$,  can be written as the superposition of two orthonormal  states $\ket +$ and $\ket -$, with complex coefficients $c_+$ and $c_-$:
\begin{equation}
\label{two_state}
  \ket{\psi}=c_+\ket{+}+c_-\ket{-}.
\end{equation}
This state vector contains all the information about the quantum system and satisfies the normalization condition $\bracket{\psi}{\psi}=1$, implying that $|c_+|^2+|c_-|^2=1$.
Here the complex coefficients are called probability amplitudes.

On the other hand, in order to perform measurements on a quantum system, it is necessary to introduce a set of  operators acting on the quantum states described by Equation \eqref{two_state}.
These operators   can be written as linear combinations of the following   $2\times 2$ Hermitian matrices. 
\begin{equation}
\label{eq:gens}
\hat\sigma_0=
\begin{pmatrix}
   1&0\\
  0&1 
 \end{pmatrix},\quad
 \hat\sigma_1=
\begin{pmatrix}
   0&1\\
  1&0 
 \end{pmatrix},\quad  \hat\sigma_2=
\begin{pmatrix}
   0&-i\\
  i&0 
 \end{pmatrix},\quad  \hat\sigma_3=
\begin{pmatrix}
   1&0\\
  0&-1 
 \end{pmatrix}
\end{equation}
Here, the identity matrix $\hat\sigma_0$ together with  the Pauli matrices $\hat \sigma_1,\hat \sigma_2$, and $\hat \sigma_3$   satisfy the well-known relation
\begin{equation}
    \hat\sigma_l\hat\sigma_m=\hat\sigma_0\delta_{lm}+i\epsilon_{lmn}\hat\sigma_n,
\end{equation}
which is the matrix representation of \eqref{eq:elm}.
Consequently, any   operator can be written as the sum
\begin{equation}
\label{herm}
\hat H = h_0\hat\sigma_0+h_1\hat\sigma_1+h_2\hat\sigma_2+h_3\hat\sigma_3.
\end{equation}
In addition, since $\hat H$ is a Hermitian matrix, the coeficients $h_k$, $(k=0,\ldots,3)$ must be real numbers.

Finally, the dynamics of the state vector is described by the Schr\"odinger equation
\begin{equation}
\label{schrod}
  i\hbar\frac{\partial \ket{\psi}}{\partial t}=\hat H\ket{\psi},
\end{equation}
where $\hat H$ represents the Hamiltonian of the TSS.
If this Hamiltonian  is time independent, then  Equation~\eqref{schrod} shows that the state of the system at any time $t$ can be obtained  by the evolution operator $\hat U=~\exp(-i\hat H t/\hbar)$ acting on the initial state $\ket{\psi_0}$, i.e.
\begin{equation}
\label{evolution}
\ket{\psi(t)}=\hat  U\ket{\psi_0}.
\end{equation}

% ------------------------------------------------------------------------
% ------------------------------------------------------------------------

\begin{thebibliography}{111}
\bibitem{Hiley2012192}
B.J. Hiley and R.E. Callaghan.
\newblock Clifford algebras and the Dirac--Bohm quantum Hamilton--Jacobi
  equation.
\newblock { Foundations of Physics}, 42(1):192--208, (2012).

\bibitem{lounesto2001clifford}
P. Lounesto.
\newblock { Clifford Algebras and Spinors}.
\newblock Cambridge University Press, (2001).

\bibitem{vaz2016introduction}
J.~Vaz and R.~da~Rocha.
\newblock { An Introduction to Clifford Algebras and Spinors}.
\newblock Oxford University Press, (2016).

\bibitem{Zou2009147}
Y.M. Zou.
\newblock Ideal structure of Clifford algebras.
\newblock { Advances in Applied Clifford Algebras}, 19(1):147--153, (2009).
  
\bibitem{Arthur2011}
J.W. Arthur.
\newblock {Understanding Geometric Algebra for Electromagnetic Theory}.
\newblock {John Wiley \& Sons, Hoboken, New Jersey}
\newblock (2011).

\bibitem{Dressel20151}
J.~Dressel, K.Y. Bliokh, and F.~Nori.
\newblock Spacetime algebra as a powerful tool for electromagnetism.
\newblock { Physics Reports}, 589:1--71, (2015).

\bibitem{Hestenes2003691}
D.~Hestenes.
\newblock Spacetime physics with geometric algebra.
\newblock { American Journal of Physics}, 71(7):691--714, (2003).

\bibitem{Lasenby1998487}
A.~Lasenby, C.~Doran, and S.~Gull.
\newblock Gravity, gauge theories and geometric algebra.
\newblock { Philosophical Transactions of the Royal Society A: Mathematical,
  Physical and Engineering Sciences}, 356(1737):487--582, (1998).

\bibitem{Lasenby2019}
A.N. Lasenby.
\newblock Geometric algebra, gravity and gravitational waves.
\newblock { Advances in Applied Clifford Algebras}, 29(4), (2019).

\bibitem{Dargys2017241}
A.~Dargys and A.~Acus.
\newblock Calculation of quantum eigens with geometrical algebra rotors.
\newblock { Advances in Applied Clifford Algebras}, 27(1):241--253, (2017).

\bibitem{doran_lasenby_2003}
C. Doran, and A. Lasenby.
\newblock { Geometric Algebra for Physicists}.
\newblock Cambridge University Press, (2003).

\bibitem{Francis2005383}
M.R. Francis and A.~Kosowsky.
\newblock The construction of spinors in geometric algebra.
\newblock { Annals of Physics}, 317(2):383--409, (2005).

\bibitem{Baylis2010517}
W.E. Baylis, R.~Cabrera, and J.D. Keselica.
\newblock Quantum/classical interface: Classical geometric origin of fermion
  spin.
\newblock { Advances in Applied Clifford Algebras}, 20(3--4):517--545, (2010).

\bibitem{McKenzie2015}
C. McKenzie.
\newblock { An Interpretation of Relativistic Spin Entanglement Using
  Geometric Algebra}.
\newblock (2015).
\newblock {Electronic Theses and Dissertations. 5652. 
\url{https://scholar.uwindsor.ca/etd/5652}}

\bibitem{Lounesto1987}
P. Lounesto and G.~P. Wene.
\newblock Idempotent structure of Clifford algebras.
\newblock { Acta Applicandae Mathematica}, 9(3):165--173, Jul (1987).

\bibitem{Morais20141}
J.P. Morais, S.~Georgiev, and W.~Sprößig.
\newblock { Real Quaternionic Calculus Handbook}.
\newblock {Birkhäuser, Basel.}
\newblock (2014).

\bibitem{baylis2004electrodynamics}
W.~Baylis.
\newblock { Electrodynamics: A Modern Geometric Approach}.
\newblock Progress in Mathematical Physics. Birkh{\"a}user Boston, (2004).

\bibitem{PhysRevLett.80.5243}
C. Bender and S. Boettcher.
\newblock Real spectra in non-Hermitian Hamiltonians having
  $\mathscr{P}\mathscr{T}$ symmetry.
\newblock { Phys. Rev. Lett.}, 80:5243--5246, Jun (1998).

\bibitem{sakurai2011modern}
J.J. Sakurai and J.~Napolitano.
\newblock { Modern Quantum Mechanics}.
\newblock Addison-Wesley, (2011).

\bibitem{cohen1991quantum}
C.~Cohen-Tannoudji, B.~Diu, and F.~Laloe.
\newblock { Quantum Mechanics}.
\newblock Wiley, (1991).

\bibitem{Hestenes2003104}
D.~Hestenes.
\newblock Oersted medal lecture 2002: Reforming the mathematical language of
  physics.
\newblock { American Journal of Physics}, 71(2):104--121, (2003).
%%%%%%%%%%%%%%%%%%%%%%%%%%%%%%%%%%%%%%%%%%%%%%%%%%%%%%%%%%%%%%%%%%%%%%%%%%%%%
\end{thebibliography}
\end{document}